\documentclass[11pt]{article}
\usepackage{color}
\usepackage{graphicx,amssymb,amsfonts,amsmath,amssymb,amscd,amstext, mathrsfs}
\usepackage{xcolor}
\usepackage{graphicx} 
\usepackage[colorlinks=true,citecolor=magenta, linkcolor=black]{hyperref}
\usepackage{amsmath}
\usepackage{empheq}
 
\newlength\dlf  

\def\be{\begin{eqnarray}}
\def\ee{\end{eqnarray}}

\def\tr{\mathop{\mathrm{tr}}}

\newcommand{\oD}{\mathring{{\rm \nabla}}}

\newcommand{\oR}{\mathring{R}}

\def \bea {\begin{equation}}
\def \eea {\end{equation}}
\def \nn {\nonumber}
\def \ts{ \textstyle}
\def \si{\sigma}
\def \de{\delta}

\def \la {\langle}
\def \ra {\rangle}
\def \rr {\raise.35ex\hbox{\small $\prime$}\kern-.17em{\mbox{\large $\imath$}}}
\def \del {\partial}
\def \dels {\partial\kern-.5em / \kern.5em}
\def \As {{A\kern-.5em / \kern.5em}}
\def \Ds {D\kern-.7em / \kern.5em}

\def \a {\alpha}

\def \b {\beta}

\def \m {\mu}
\def \n {\nu}

\def\frac#1#2{{#1\over #2}}

\begin{document}

\thispagestyle{empty}
\begin{flushright} 
YITP-SB-16-08 
\end{flushright}

\begin{center} 
\vspace{2.5cm} {\LARGE{
Boundary Anomalies and Correlation Functions}}\\
\vspace{0.7 cm}  Kuo-Wei Huang \\ 
\vspace{0.6 cm} {\it Department of Physics and Astronomy,\ \\ 
C.~N.~Yang Institute
for Theoretical Physics,\\
Stony Brook University, Stony Brook, NY  11794}

\vspace{0.8cm}

\end{center}

\begin{abstract} \noindent 
It was shown recently that boundary terms of
conformal anomalies recover the universal contribution to the
entanglement entropy and also play an important role in the boundary
monotonicity theorem of odd-dimensional quantum field theories.
Motivated by these results, we investigate relationships between 
boundary anomalies and the stress tensor correlation functions in
conformal field theories. In particular, we focus on how the conformal
Ward identity and the renormalization group equation are modified by
boundary central charges. 
Renormalized stress tensors induced by boundary Weyl invariants are also discussed, with examples in spherical and cylindrical geometries.

\end{abstract}

\newpage

\tableofcontents

\section{Introduction}

As a non-trivial extension of Poincar\'e symmetry, Weyl invariance
imposes significant constraints on the structure of correlation
functions. For a generally non-conformal quantum field theory (QFT), the
correlation functions approach those of a conformal field theory ($\rm
CFT$) sitting at one end point of the renormalization group (RG) flow
at short distance, and those of a $\rm CFT$ sitting at another end
point of the RG flow at long distance.  It is therefore of great
importance to search for a general principle constraining the flows
between the two end points. In $d=2$, the well-known c-theorem was proved in
\cite{Zamolodchikov1986} thirty years ago.  Conjectured first by
\cite{Cardy1988}, a proof of the a-theorem in $d=4$ was given only
recently by \cite{Komargodski:2011vj}  \cite{Komargodski:2011xv}, using
the so-called dilaton anomaly effective action.  The proof has not been
found for $d=6$ QFTs. See \cite{Elvang:2012st} \cite{Cordova:2015fha}
\cite{Heckman:2015axa} \cite{Grinstein:2014xba} for the recent progress.
 In proving these monotonicity theorems, the central charges, defined as
the coefficients of the trace anomaly when embedding the theory in a
curved background, play the central role.  In particular, it is the
central charge of the topological Euler density that satisfies the
irreversibility of the RG flow.

There is no Weyl anomaly in odd-dimensional (compact) manifolds, and
hence a definition for central charge becomes elusive.  An alternative
candidate that satisfies the monotonic behaviour along the RG flow in
$d=3$ was suggested by \cite{Jafferis:2010un} as the Euclidean path
integral of the CFT conformally mapped to $S^3$.  See
\cite{Jafferis:2011zi} \cite{Klebanov:2011gs} for further discussions on
such an F-theorem; \cite{Giombi:2014xxa} points out an interpolation  
between the a-theorem and the F-theorem at the fixed points.  The universal part of the
partition function on $S^3$ can be further identified as the constant
piece of the vacuum entanglement entropy (EE) across a disk. Moreover,
with the help of strong subadditivity \cite{SSA}, the
irreversibility of the RG flow in $d=3$ can be proved
\cite{Casini:2012ei}.  It remains  an important question whether one can
link the fundamental properties of EE to the monotonicity theorem for
spacetime dimensions higher than three.  (See \cite{Myers:2010tj}
\cite{Myers:2010xs}  for examining the general RG flow using 
holography.)

However, in many physical systems, in particular in condensed matter
physics, edge effects are important.  It has also been shown recently that the
universal part of EE can be understood purely as a boundary effect
\cite{Herzog:2015ioa} (so in some sense the ``area/boundary law" of EE
is extended to also include the UV cut-off-independent log-term), which
solves the main puzzle left from an earlier attempt
\cite{Casini:2011kv}. The study of EE in $d=3$ with a spacetime boundary
is discussed in a more recent work \cite{Fursaev:2016inw}. Most
important for this paper is the fact that there are additional boundary invariants in the presence of a boundary, for any-dimensional CFTs.  
To make a connection with the monotonicity theorem,
we notice that the boundary central charge  orders the boundary RG flow
in $d=3$ \cite{Jensen:2015swa};  this b-theorem is then
a generalization of the boundary g-theorem \cite{g}. See
\cite{Nozaki:2012qd} for a related discussion.

Motivated by these results, we here consider how the boundary central
charges affect the conformal Ward identities and the RG equation of
stress tensor correlation functions. We will largely focus on $d=4$, but
will also discuss $d=3$ where only  boundary anomalies exist. (In
$d=2$, there is no new boundary anomaly and only a boundary term of
the Euler density is needed, which we discuss briefly in the appendix.)

To set the stage for our discussion, in the next section we first review
the conformal invariance and the correlation function. The explicit
expression for the stress tensor three-point function is rather bulky
and we will refer the reader to \cite{Osborn:1993cr}
\cite{Erdmenger:1996yc}. Our main focus here instead is the additional
contribution to correlation functions from the local counter-terms
(conformal anomaly), in particular when a manifold has a boundary.  In
Sec.\ 3, we revisit several main identities of correlation functions. We
are interested in correlation functions in flat bulk spacetime but we
allow a generally curved, codimension-1 boundary. The boundary will be
assumed to be compact and smooth (no corners). When revisiting these
identities, we  will not set the Weyl anomaly to be zero in the flat
limit since the boundary terms survive even in the flat space.
The boundary local counter-terms also contribute non-trivially to
the stress tensor in the flat limit, so we will not drop the stress
tensor either. 
These identities then generalize the
ones given in the literature (for example, \cite{Osborn:1993cr}
\cite{Erdmenger:1996yc}).

The $d=4$ {\rm CFTs} will be considered in Sec.\ 4. We first discuss the
RG equation of the three-point function and the conformal Ward identity
for a compact manifold. Then, we generalize these results by introducing
a boundary. It is found that the RG equation is modified by a special
boundary central charge defined by a Weyl invariant constructed solely
from the extrinsic curvature. The Ward identity is also modified due to
the boundary counter-terms. Moreover, we obtain stress tensors in the vicinity of a boundary whose values are determined by boundary central charges.  Examples in ball and cylindrical  geometries are given.  A similar analysis for $d=3$ {\rm CFTs} is in Sec.\ 5, where the story becomes simpler because there is no bulk conformal anomaly and
there are only two simple Weyl anomalies living on the boundary.

Among our new results, the most important ones are: a formal expression of the  trace conformal Ward identity for general $d$ \eqref{id3}; an RG equation for the three-point function  \eqref{rg4b}, the Ward identity \eqref{cb},  a $b_1$-type stress tensor for a 4-cylinder \eqref{rst}, $a$-type stress tensors for a 4-ball \eqref{abt} and for a 4-cylinder \eqref{st2} for $d=4$; the Ward identity \eqref{3dw}, an RG equation for the two-point function \eqref{3rg}, a $c$-type stress tensor for a 3-cylinder \eqref{rst3}, $a$-type stress tensors for a 3-ball \eqref{a3b} and for a 3-cylinder \eqref{a3c} for $d=3$.  

In the conclusion we point out related questions. The appendix contains useful formulae for metric variation, and also a brief discussion on $d=2$ {\rm CFTs} with a boundary.

\section{Conformal Invariance and Anomalous Terms}

For a conformal transformation $g$, $x_\mu \to x'_{\mu}=(g x)_\mu$, we
can define a local orthogonal matrix on $\Bbb { R}^{d-1,1}$ 
\be
{\rm R}^{\mu}_{~\nu}(x)= \Omega^g (x) {\partial x'^\mu \over \partial x^\nu}\ ,~~ {\rm R}^{\mu}_{~\a}(x) {\rm
R}^{\nu}_{~\b}(x)\eta_{\m\n}=\eta_{\a\b} \ , 
\ee 
where $\Omega^g (x)$ is
the scale factor showing up in the transformed line element $ds'^2=
\Omega^g (x)^{-2} ds^2$. In the Euclidean signature we replace
$\eta_{\m\n}$ by $\delta_{\m\nu}$ on $\Bbb { R}^{d}$ and ${\rm
R}_{\mu\nu}(x)$ becomes a local rotation matrix belonging to the group
$O(d)$. One can generate a conformal transformation by a combination
of translations, rotations and inversions, although the conformal group
$ SO(d+1,1)$ can only be formed by even numbers of inversions since
the inversion itself is not connected to the identity. If we consider an
inversion through the origin, $x'_\mu= {x_\mu \over x^2}$, we have
$\Omega(x)=x^2$ and 
\be
{\rm R}_{\mu \nu}(x)= I_{\mu\nu}(x)=
\delta_{\mu\nu}-2 {x_\mu x_\nu \over x^2} , ~~ I^{\mu\nu}(x'-y')={\rm
R}^{\mu}_{~\a}(x){\rm R}^{\nu}_{~\b}(y) I^{\a \b}(x-y) \ . 
\ee 
The
matrix $I_{\mu\nu}(x-y)$ transforms like a vector and it can be regarded
as a parallel transport for the conformal transformations.  For three
points, $x,y,z,$ one can also  define a covariant vector that transforms
homogeneously. At the point $z$ such a vector is given by
\be 
Z_\mu= {1\over 2}
\partial^{(z)}_\mu \ln{(z-y)^2\over (z-x)^2}={(x-z)_\mu \over (x-z)^2}-
(x\leftrightarrow y) \ ,~ Z'_{\mu}= \Omega(z) {\rm
R}_{~\mu}^{\lambda}(z) Z_\lambda \ . 
\ee 
Similarly, at points $x$, $y$ one
can define covariant vectors $X_\mu$ and $Y_\nu$ via cyclic permutation.

Denote a conformal primary operator as ${\cal O}^i(x)$ with $i$
representing components in some representation of the rotation group. To
construct the correlation functions, it is useful to adopt the induced
representations \cite{Salam} to write the conformal transformation as 
\be
\label{pt} {\cal O}^i(x) \to {\cal O'}^i(x')= \Omega(x)^\Delta
D^i_j({\rm R}(x)) {\cal O}^j(x) \ , 
\ee 
where $\Delta$ is the conformal
dimension; $D^i_j({\rm R}(x))$ is the matrix in the associated representation.
The conjugate representation of \eqref{pt} is ${\cal \bar O'}_i(x')=
\Omega(x)^\Delta {\cal \bar O}_j(x) (D({\rm R}(x))^{-1})^j_i$, where $ {\cal\bar
O}_i(x)$ stands for the conjugate field. The two-point function for
conformal primary operators can be written by 
\be 
\label{oo} \la
{\cal O}^i (x){\cal \bar O}_j(y) \ra= {C_{\cal O}\over (x-y)^{2\Delta} }
D^i_j(I(x-y)) \ , 
\ee 
in the irreducible representations of $ O(d)$.
The overall constant $C_{\cal O}$ might be set to one by operator
redefinition.

Our main subject of interest is the correlation functions of the stress
tensor $T_{\m\n}$, which  is a symmetric tensor satisfying $\partial^\mu
T_{\m\n}=0$ and $T^\mu_\mu=0$. The latter property implies conformal
invariance at the classical level.  $T_{\m\n}$ has the scale dimension $d$ required also by the conformal invariance. 
Now applying \eqref{oo} to the stress tensor gives 
\be 
\label{contact} 
\la T_{\mu\n}(x)T_{\a\b}(y)\ra= {C_T \over (x-y)^{2d}} {\cal I}_{\m\n,\a\b}(x-y) \ , 
\ee
where $C_T$ is a constant coefficient and 
\be 
&&{\cal I}_{\mu\n,\a\b}(x-y)=  I_{\m \lambda}(x-y) I_{\nu\rho}(x-y)
{L}^{\lambda\rho}_{~~\a\b} \ , ~~ 
{L}^{\lambda\rho}_{~~\a\b} \equiv
{1\over 2}\Big(\delta^\lambda_\a \delta^\rho_{\b}+\delta^\rho_\a
\delta^\lambda_\b- {2\over d} \delta_{\a\b}\delta^{\lambda\rho}\Big) \ .
\ee 
Here $L^{\lambda\rho}_{~~\a\b}$ is a projection operator onto the
space of a symmetric, traceless tensor. The expression of the
three-point function becomes rather involved  \cite{Osborn:1993cr}
\cite{Erdmenger:1996yc}. To keep the expression simple we will not list
its full expression but  simply write it as 
\be 
\label{ttt3} 
\la T_{\mu\n}(x) T_{\sigma\rho}(y) T_{\a\b}(z)\ra=\Gamma_{\m\n\sigma\rho\a\b}(x,y,z) \ . 
\ee 
The structure of $\Gamma_{\m\n\a\b\sigma\rho}(x,y,z)$ is determined by the conformal
symmetry and the stress tensor conservation. Its explicit form will not be too
relevant in the present discussion. However, it is important to know
that there are three independent forms (in $d=3$ there are two and only
one in $d=2$)  for the stress tensor three-point function
\cite{Osborn:1993cr}.

Here we focus on the additional contribution to \eqref{ttt3} coming from
possible local counter-terms in the effective action. These
counter-terms give precisely the Weyl anomaly.  We will adopt
dimensional regularization working in $d\to d+\epsilon$ dimensions and refer to these counter-terms as the anomaly effective action denoted by
$\widetilde W\equiv {\mu^{\epsilon}\over \epsilon} \tilde W$.  By
including possible counter-terms we have 
\be 
\label{ttt2} 
\la T_{\mu\nu}(x) T_{\sigma\rho}(y) T_{\alpha\beta}(z)\ra=\Gamma_{\mu\nu\sigma\rho \alpha\beta}(x,y,z)+ (-2)^3 {\delta^3 \over 
 \delta g^{\mu\nu}(x) g^{\sigma\rho}(y) \delta g^{\alpha\beta}(z)} \widetilde W \ .
\ee 
In $d=4$, there are two types of local counter-terms: one type-A
anomaly and one type-B anomaly. (We will remove the scheme-dependent type-D total derivative anomaly.) The contributions from the bulk anomalies were first calculated by \cite{Osborn:1993cr}
and we will review them. Notice that the scale $\mu$-dependent part in
the second term of \eqref{ttt2} determines the RG equation of the
three-point function. 

Our main input 
is the new correction from the boundary terms of the
effective action, i.e. the second term in 
\be 
\widetilde W={\mu^{\epsilon}\over \epsilon }\Big( \int_{\cal M} \tilde W+
\int_{\partial {\cal M}}\tilde W_{\rm bry} \Big) \ . 
\ee 
The effective
action must reproduce the boundary anomalies via the conformal
transformation.   As we will discuss in more detail, 
in addition to the Euler characteristic boundary term needed to preserve topological invariance, there exists new boundary Weyl
invariants. To our knowledge, the corrections to the stress tensor
correlation functions from the boundary anomalies have not been
discussed in the literature.

\section{Ward Identities and Anomalies Revisited}

Considering a field theory coupled to a non-dynamical, curved
background, we define the stress tensor in the Euclidean signature by
\be 
\label{stress tensor} 
\la T_{\mu\nu}(x) \ra= -{2\over \sqrt{g}}
{\delta  {{W}} \over \delta g^{\mu\nu}(x)} \ , 
\ee 
where the
functional ${{W}}\equiv {{W}}(g(x))$ is the effective action (the
generating functional for connected Green's functions).  The classical
stress tensor trace vanishes for CFTs but at the quantum level, the
regularization and renormalization introduce conformal symmetry breaking
counter-terms that result in $\la T^\mu_\mu (x) \ra\neq 0$; $\widetilde
W$ denoted earlier is a part of $ W$.  We assume the theory is regulated in a diffeomorphism-invariant way and we will focus on the vacuum.
The trace anomaly is a function of intrinsic curvatures (for a compact
manifold) that must vanish in the flat limit.  However, the flat space
correlation functions depend on central charges: a result reminding us to
adopt the correct order of limits, that the flat limit is imposed only
after performing the metric variation.

We have the following definitions of the functional differentiation: 
\be
\label{tt} \la T_{\mu\nu} (x) T_{\sigma\rho}(y) \ra &=&
\lim_{g_{\mu\nu}\to \delta_{\mu\nu}}\Big((-2)^2 {\delta^2 \over \delta
g^{\sigma\rho}(y)\delta g^{\mu\nu}(x) } W|_{g=0} \Big) \ , \\
\label{ttt} \la T_{\mu\nu} (x) T_{\sigma\rho}(y) T_{\a\b} (z)\ra &=&
\lim_{g_{\mu\nu}\to \delta_{\mu\nu}}\Big((-2)^3 {\delta^3 \over \delta
g^{\a\b} (z)\delta g^{\sigma\rho}(y)\delta g^{\mu\nu}(x) } W|_{g=0}
\Big) \ , 
\ee 
where $W|_{g=0}$ is the standard procedure getting rid of
the source term in the action after performing the variation. Notice that we have
restricted the results to flat space at the end of the computation. From
these definitions, we can derive the following identity: 
\be 
\label{id1}
&&\la T_{\mu\nu} (x) T_{\sigma\rho}(y) T_{\a\b}
(z)\ra \nn\\ 
&=&-8\lim_{g_{\mu\nu}\to
\delta_{\mu\nu}}{\delta \over \delta g^{\a\b}(z)} {1\over
{\sqrt{g(y)}}}{\delta \over \delta g^{\sigma\rho}(y)}{1\over
{\sqrt{g(x)}}}{\delta \over \delta g^{\mu\nu}(x)}W|_{g=0}+
B_{\mu\nu\sigma\rho\a\b}(x,y,z)\nn\\ 
&&+{\delta_{\a\b}
\Big(\delta^d(y-z)}+\delta^d(x-z)\Big) \la
{T_{\mu\nu}(x)T_{\sigma\rho}(y)}\ra +{\delta_{\sigma\rho} \delta^d(x-y)}
\la T_{\mu\nu}(x){T_{\a\b}(z)}\ra \ , 
\ee 
where 
\be
B_{\mu\nu\sigma\rho\a\b}(x,y,z) = \Big(\delta_{\sigma\a}
\delta_{\rho\b}+\delta_{\sigma\b} \delta_{\rho\a}-{
2}{\delta_{\a\b}}{\delta_{\sigma\rho}\Big) \delta^d(x-y) \delta^d(y-z)
\la T_{\mu\nu}(x)\ra|_{\delta_{\mu\nu}}} 
\ee 
vanishes for a manifold
without a boundary because the expectation value of the stress tensor
vanishes in flat space. 
However, we will find a non-trivial contribution to $\la
T_{\mu\nu}(x)\ra|_{\delta_{\mu\nu}}$ when we include boundary Weyl
anomalies.  The conservation of the stress tensor implies 
\be 
\label{id2}
&&\partial^{\m} \la T_{\mu\nu} (x) T_{\sigma\rho}(y) T_{\a\b}
(z)\ra\nn\\ &=&\partial_{\nu}
\Big(\delta^d(x-y)+\delta^d(x-z)\Big) \la
{T_{\sigma\rho}(y)T_{\a\b}(z)}\ra +\partial^{\m}
B_{\mu\nu\sigma\rho\a\b}(x,y,z)\nn\\ 
&&+\Big\{\partial_\sigma
\Big(\delta^d(x-y) \la {T_{\rho\nu}(x) T_{\a\b}(z)}\ra\Big) +\sigma
\leftrightarrow \rho\Big\} +\Big\{ \partial_\a \Big(\delta^d(x-z) \la
{T_{\n\b}(x)T_{\sigma\rho}(y)}\ra \Big) +\a \leftrightarrow \b\Big\} \ . \nn\\ 
\ee

An important relation that we can also derive from these definitions is
the following trace conformal Ward identity relating the three-point
function and the two-point function: 
\be 
\label{id3} &&\la T_{\mu}^{\mu}
(x) T_{\sigma\rho}(y) T_{\a\b} (z)\ra\nn\\
&=& 2  \Big(\delta^d(x-y)+\delta^d(z-x) \Big)\la T_{\sigma\rho}(y) 
T_{\a\b} (z) \ra+ 4 \lim_{g_{\mu\nu}\to \delta_{\mu\nu}} \Big({\delta^2
\over \delta g^{\a\b} (z)\delta g^{\sigma\rho}(y)}  \la T^\mu_\mu(x)
\ra\Big) \nn\\ 
&&-2\lim_{g_{\mu\nu}\to \delta_{\mu\nu}}
\Big({\delta_{\a\b}}\delta^d(x-z)  {\delta \over \delta
g^{\sigma\rho}(y)}+ {\delta_{\sigma\rho}}\delta^d(x-y)  {\delta \over
\delta g^{\a\b}(z)}\Big)\la T^{\mu}_{\mu}(x) \ra\nn\\
&&+S_{\a\b\sigma\rho}(x,y,z)\la T^\m_\m(x) \ra|_{\delta_{\m\n}}\ , 
\ee 
where 
\be
\label{s}
S_{\a\b\sigma\rho}(x,y,z)&=& \Big(
\delta_{\a\sigma} \delta_{\b\rho}+\delta_{\a\rho} \delta_{\b\sigma}+\delta_{\a\b} \delta_{\rho\sigma} \Big)\delta^d(x-y)\delta^d(x-z) \ . 
\ee 
For CFTs in a compact
manifold the last two lines of \eqref{id3} do not contribute, and therefore these terms are not included in the literature. 
(Take $d=4$ CFTs for example. After performing a metric
variation on the stress tensor trace, the result vanishes in the flat limit since the trace anomaly is a function of curvature squared.  The same argument applies for higher dimensions.) 
We see that in general the existence of the boundary Weyl
anomaly modifies the trace Ward identity.

\section{Four-Dimensional CFTs} 
\subsection{Compact Manifold}

In this section we reproduce several main results for $d=4$ CFTs
without boundary terms.  We follow \cite{Osborn:1993cr}
\cite{Erdmenger:1996yc} closely but the way we adopt the dimensional
regularization will be slightly different in the details.

The conformal anomaly reads 
\be 
\label{4dtraceb} 
\la T^\mu_\mu (x)\ra= {1\over 16 \pi^2} \Big( c W_{\mu\si\rho\nu}^2-a E_4+ {\gamma} \Box R
\Big) \ . 
\ee 
The type-A anomaly is defined by the Euler density, 
\be
E_4 ={1\over 4} \delta^{\a\b\gamma\delta}_{\mu\nu\sigma\rho}
R^{\mu\nu}_{~~\a\b} R^{\sigma\rho}_{~~\gamma\delta}
=R^2_{\mu\nu\sigma\rho}-4 R^2_{\m\n}+R^2 \ , 
\ee 
where $\delta^{\a\b\gamma\delta}_{\mu\nu\sigma\rho}$ is the fully
anti-symmetrized product of four Kronecker delta functions. The Euler
number $\chi$ is given by $\int_{\cal M} d^4 x \sqrt{g} E_4= 4 \pi
\chi$. The type-B anomaly defines the central charge $c$; the Weyl
tensor in d-dimensions (for $d>3$) is given by 
\be 
\label{weyl}
W^{(d)}_{\mu\si\rho\nu} = R_{\mu\si\rho\nu} - {2\over d-2} \Big(
g_{\mu[\rho}R_{\nu]\si} - g_{\si[\rho}R_{\nu]\mu}  -
{g_{\mu[\rho}g_{\nu]\si}\over (d-1)}R\Big) \ , 
\ee 
which has basic
properties: $W_{\mu\si\rho\nu}=W_{[\mu\si][\rho\nu]}$,
$W_{\mu[\si\rho\nu]}=0$ and $W^{\mu}_{~\si\rho\mu}=0$.  The last term in
\eqref{4dtraceb}, referred to as the type-D anomaly, is scheme-dependent:
adding an $R^2$ local counter-term shifts its anomaly coefficient. Here
we use this freedom to remove this anomaly. See \cite{Herzog:2013ed}
and \cite{Huang:2013lhw} for related discussions on the type-D
anomaly.\footnote{It is shown recently that this type-D
anomaly can be intrinsically fixed by supersymmetry \cite{Assel:2015nca}.}

Following expression \eqref{ttt2}, we write the stress tensor
three-point function as 
\be 
\la T_{\mu\nu}(x) T_{\sigma\rho}(y)
T_{\alpha\beta}(z) \ra= \Gamma_{\mu\nu\sigma\rho \alpha\beta}(x,y,z)- 
{\mu^{\epsilon} \over 2 \pi^2 \epsilon}\Big(c D_{\mu\nu\sigma\rho \alpha\beta}
^W(x,y,z)-a D_{\mu\nu\sigma\rho
\alpha\beta} ^E(x,y,z)\Big)\ , 
\ee 
where the $\mu$-dependent part is determined by
the counter-terms. We take 
\be 
\label{DE}
D_{\mu\nu\sigma\rho\alpha\beta}^E(x,y,z)&=& \lim_{g_{\mu\nu}\to \delta_{\mu\nu}} { \delta^3\over  ~\delta{g^{\mu\nu}(x) \delta g^{\sigma\rho}(y) \delta
g^{\a\b}(z)}} \int_{{\cal M}} d^d x' \sqrt{g} E_{4}(x') \ , \\
\label{DW}
D_{\mu\nu\sigma\rho\alpha\beta}^W(x,y,z)&=&\lim_{g_{\mu\nu}\to
\delta_{\mu\nu}} { \delta^3 \over   ~\delta{g^{\mu\nu}(x) \delta
g^{\sigma\rho}(y) \delta g^{\a\b}(z)}} \int_{{\cal M}} d^d x' \sqrt{g}
W^{2(d)}_{\mu\nu\lambda\rho}(x') \ . 
\ee  In the dimensional regularization scheme, we consider a different treatment on the
totally-antisymmetic  tensor compared to \cite{Osborn:1993cr}
\cite{Erdmenger:1996yc}. We write the Euler density as
\be 
\label{int} 
\int_{{\cal M}} d^d x  \sqrt{g} E_4 = \int_{{\cal M}}
\frac{ \left( \bigwedge_{j=1}^d dx^{\mu_j} \right) }{4(d-4)!} {R^{a_1
a_2}}_{\mu_1 \mu_2} {R^{a_{3} a_4}}_{\mu_{3} \mu_4} e^{a_{5}}_{\mu_{5}}
\cdots e^{a_d}_{\mu_d} \epsilon_{a_1 \cdots a_d} \ . 
\ee 
Defining
$\sigma(x)$ as the Weyl transformation parameter, $g_{\mu\nu}(x)\to
e^{2\sigma} g_{\mu\nu}(x)$, we have  
\be \lim_{d\to 4} {\delta\over (d-4) \delta \sigma(x)} \int_{{\cal M}} d^d x'\sqrt{g} \Big({E_4(x')},W^{2(d)}_{\mu\nu\lambda\rho}(x')\Big)=\sqrt{g}\Big( {E_4(x)},W^{2}_{\mu\nu\lambda\rho}(x)\Big) \ ,
\ee   ($W^{2}_{\mu\nu\lambda\rho}\equiv W^{2(d=4)}_{\mu\nu\lambda\rho}$)
which confirms that the counter-terms produce the Weyl anomaly. 

Let us first consider a metric variation on \eqref{int}.
Performing integration by parts and recalling the metricity and the
Bianchi identity, in varying the integrated Euler density we in fact
only need to vary the vielbeins.  From the relation $ 2\delta / \delta
g^\nu_\mu = e_{(\nu}^a \delta / \delta e^a_{\mu)}$, we obtain 
\be
\label{firstvg} \frac{\delta}{\delta g_\mu^\nu(x)} \int_{{\cal M}} d^d
x' \sqrt{g} E_4 = \frac{\sqrt{ g}}8 {R^{\nu_1 \nu_2}}_{\mu_1 \mu_2}
{R^{\nu_{4} \nu_4}}_{\mu_{3} \mu_4} \,  \delta^{\mu_1 \cdots \mu_4
\mu}_{\nu_1 \cdots \nu_4 \nu}\ . 
\ee 
After performing two more metric
variations and imposing the flat space limit, the result is 
\be
\label{new4da} 
D_{\mu\nu \sigma\rho \alpha\beta}^E(x,y,z)= -{1\over 4}~\Big(\epsilon_{\m\sigma\a \eta \xi} \epsilon_{\n\rho\beta \zeta\delta}\partial^{\eta}
\partial^\zeta \delta^4(x-y)\partial^{\xi}\partial^\delta \delta^4(x-z)+ \sigma \leftrightarrow \rho, \a \leftrightarrow \b \Big) \ .
 \ee  
However, the a-charge does not contribute to the
RG equation of the three-point function because $\lim_{d\to
4}D_{\mu\nu\sigma\rho \alpha\beta}^E(x,y,z)=0$ due to the five totally
anti-symmetized indices. It is the c-charge that controls the RG
equation 
\be 
\label{rge4d} 
\mu {\partial\over \partial \mu} \la
T_{\mu\nu}(x) T_{\sigma\rho}(y) T_{\alpha\beta}(z) \ra = -{c\over
2\pi^2} \lim_{d\to 4} D^W_{\mu\nu\sigma\rho\a\b} (x,y,z) \ . 
\ee 
While the explicit expression of $D^W_{\mu\nu\sigma\rho\a\b} (x,y,z)$ is not
essential for our purpose, we remark that, from the following
re-writing 
\be 
\label{rewq}
W^{2(d)}_{\mu\nu\lambda\rho} = E_4+{4(d-3)\over (d-2)}
Q_d,~~~Q_d= R^2_{\mu\nu}-{d\over 4(d-1)} R^2 \ , 
\ee 
one can also say the RG flow is dictated by the Q-curvature\footnote{In $d=4$ the notion of Q-curvature  was introduced by Branson and {\O}rsted \cite{qcurvature} as a global conformal invariant. However, note that the curvature $Q_{4}$ in \eqref{rewq} is different from the Q-curvature defined in \cite{qcurvature} by a total derivative, $ \Box R$.  For an analysis of Q-curvature in $d=6$, see \cite{q6}.}: 
\be \label{qq}
D_{\mu\nu\sigma\rho\alpha\beta}^W(x,y,z)= 2 \lim_{g_{\m\n}\to
\delta_{\m\n}} \Big({ \delta^3 \over   ~\delta{g^{\m\n}(x) \delta
g^{\sigma\rho}(y) \delta g^{\a\b}(z)}} \int_{\cal{M}} d^d x' \sqrt{g}
Q_d(x')\Big)\ . 
\ee 
The above third-order metric variation can be
performed straightforwardly. However, we do not find a compact expression so
we do not list the result here.

Finally, given the conformal anomaly \eqref{4dtraceb}, the Ward identity
\eqref{id3} can also be computed and the result is given by
\cite{Osborn:1993cr} 
\be 
\label{ward} 
&&\la T_{\mu}^{\mu} (x)
T_{\sigma\rho}(y) T_{\a\b} (z)\ra \nn\\ 
&&= 2 
\Big(\delta^4(x-y)+\delta^4(z-x) \Big)\la T_{\sigma\rho}(y)  T_{\a\b}
(z) \ra+ {1\over 4 \pi^2} \Big(c B_{\sigma\rho\a\b}(x,y,z)-a
A_{\sigma\rho\a\b}(x,y,z)\Big) \ . 
\ee 
The anomalous terms are
determined by the second-order expansion of the anomalies: 
\be 
\label{ab1}
A_{\sigma\rho\a\b}(x,y,z)&=&\lim_{g_{\mu\nu}\to \delta_{\mu\nu}}
{\delta^2 \over \delta{g^{\sigma\rho}(y) \delta g^{\a\b}(z)}} 
E_{4}(x)\nn\\
&=&{-}\Big\{\epsilon_{\sigma\alpha\lambda\mu}\epsilon_{\rho\b\delta\n}\partial^\m \partial^\n\Big(\partial^\lambda
\delta^4(x-y)\partial^\delta\delta^4(x-z)\Big)+ \sigma \leftrightarrow
\rho \Big\} \ , \\ 
\label{ab2}
B_{\sigma\rho\a\b}(x,y,z)&=&\lim_{g_{\mu\nu}\to \delta_{\mu\nu}}
{\delta^2 \over \delta{g^{\sigma\rho}(y) \delta g^{\a\b}(z)}} 
W^{2}_{\mu\nu\lambda\delta}(x) \nn\\ &=&{8} \Big(
P_{\sigma\m\n\rho,\a\gamma\delta\b}\partial^\m
\partial^\n \delta^4(x-y)\partial^\gamma\partial^\delta
\delta^4(x-z)\Big) \ , 
\ee 
where a projector is introduced that shares
the same symmetries as the Weyl tensor: 
\be 
\label{P}
P_{\mu\sigma\rho\nu,\alpha\gamma\delta\beta}&=&{\ts {1\over 12}}\big(
\de_{\mu\alpha}\de_{\nu\beta}\de_{\si\gamma}\de_{\rho\de} +
\de_{\mu\de}\de_{\si\beta}\de_{\rho\alpha}\de_{\nu\gamma} - \mu
\leftrightarrow \si, \nu \leftrightarrow \rho \big ) \nn\\ 
&& +
{\ts{1\over 24}}\big(
\de_{\mu\alpha}\de_{\nu\gamma}\de_{\rho\de}\de_{\si\beta} - \mu
\leftrightarrow \si, \nu \leftrightarrow \rho, \alpha \leftrightarrow
\gamma , \de \leftrightarrow \beta \big ) \nn\\ 
&& - {\ts {1\over
16}}\big( \de_{\mu\rho}\de_{\alpha\delta}\de_{\si\gamma}\de_{\nu\beta} +
\de_{\mu\rho}\de_{\alpha\delta}\de_{\si\beta}\de_{\nu\gamma} - \mu
\leftrightarrow \si, \nu \leftrightarrow \rho, \alpha \leftrightarrow
\gamma , \de \leftrightarrow \beta \big ) \nn\\ 
&& + {\ts {1\over
12}}\Big( \de_{\mu\rho} \de_{\nu\si} - \rho \leftrightarrow \n \big )
\big ( \de_{\alpha\de}\de_{\beta\gamma} -
\delta \leftrightarrow \b \Big) \ . 
\ee 
(To the first order in
the metric expansion, one has $W_{\mu\si\rho\nu} \sim 2
P_{\mu\si\rho\nu,\alpha\gamma\delta\beta} \del^\gamma \del^\delta \delta
g^{\alpha\beta}.$) 
From the general statement \cite{Osborn:1993cr} that
there are three independent coefficients in the three-point function,
the relation \eqref{ward} implies that there exists a linear relation
between the three coefficients and the central charges $a$ and $c$. (Assume that the type-D anomaly is removed.)

\subsection{Boundary Terms of Anomalies and Effective Action}

The complete classification of possible boundary terms based on the
Wess-Zumino consistency for $d=4$ {\rm CFTs} was carried out recently by
\cite{Herzog:2015ioa}. Denoting the induced boundary metric as
$h_{\mu\nu}=g_{\mu\nu}-n_\mu n_\nu$  where $n_\mu$ is the unit-length,
outward-pointing normal vector to $\partial {\cal M}$, the full Weyl anomaly
with boundary terms is given by 
\be 
\label{4dtrace} 
\la T^\mu_\mu (x)\ra=
{1\over 16 \pi^2} \Big( c W_{\mu\nu\lambda\rho}^2- a E_4\Big)
+{\delta(x_{\perp})\over 16 \pi^2}  \Big(a E^{\rm{(bry)}}_4+ b_1 \tr\hat{K}^3+b_2
h^{\alpha\gamma}\hat{K}^{\beta\delta}W_{\alpha\beta\gamma\delta} \Big) \ , 
\ee  
where $\delta(x_{\perp})$ is a Dirac delta function with support on the boundary.
The Chern-Simons-like boundary term of the Euler Characteristic,
\be 
\label{q4} E^{\rm{(bry)}}_4
= -4 \delta^{\mu_1\mu_2\mu_3}_{\nu_1\nu_2\nu_3}~
K^{\nu_1}_{\mu_1}\left({1\over 2} R^{\nu_2 \nu_3}{}_{\mu_2 \mu_3} +
{2\over 3} K^{\nu_2}_{\mu_2} K^{\nu_3}_{\mu_3} \right)
=4\left( 2
\mathring{E}_{\alpha\beta}K^{\alpha\beta} + \frac{2}{3}\text{tr}K^3 -
K \tr K^2+\frac{1}{3}K^3\right) \ , 
\ee 
is used
to supplement the bulk density $E_4$ to preserve the topological
invariance. (In the literature, it is also referred to as the boundary term
for the Lovelock gravity. See \cite{review} or \cite{Herzog:2015ioa} for
a review.) We have denoted $\mathring{E}_{\alpha\beta}$ as the boundary Einstein tensor. There are two additional boundary Weyl invariants in the
anomaly \eqref{4dtrace} and we refer to $b_1$ and $b_2$ as ``boundary"
central charges.   The Weyl curvature in the last term of \eqref{4dtrace} and the Riemann curvature in \eqref{q4}  are pulled-back tensors. The traceless part of the
extrinsic curvature is given by 
\be
\hat{K}_{\alpha\beta}=K_{\alpha\beta} - \frac{K}{d-1}h_{\alpha\beta} \ ,
\ee 
with $d=4$ here. (We still adopt the Greek indices for these boundary tensors, but note that their normal component is empty.)  It is an important property that $\hat{K}_{\alpha\beta}$ transforms covariantly under the Weyl scaling. The boundary $b_1$-type anomaly to our knowledge first appeared in
\cite{b1}; \cite{b2} first pointed out the boundary $b_2$-type
anomaly. Spelling out the expressions, we have 
\be 
\tr\hat{K}^3&=&\tr K^3-K \tr K^2+{2\over 9} K^3 \ , \\
h^{\alpha\gamma}\hat{K}^{\beta\delta}W_{\alpha\beta\gamma\delta} &=& 
R^{\mu}_{~\nu \lambda \rho} K_{\mu}^{\lambda} n^\nu n^\rho -{1\over 2}
R_{\mu\nu} (n^\mu n^\nu K + K^{\mu\nu}) +{1\over 6} KR  \ . 
\ee 
Notice
that, because of these boundary terms, the Weyl anomaly (the log-divergent term
of the partition function) of CFTs does not vanish even in flat space.
 In particular, the boundary $b_1$-type anomaly constructed solely from
the extrinsic curvature exists in any dimensions, and it might play a
role to order the (boundary) RG flows in any-dimensional QFTs.

The conformal anomaly has its origin tracing back to the local
counter-terms. Adopting the dimensional regularization, we should include
the corresponding boundary terms in the effective action. We have the following
identities: 
\be 
&&\lim_{d\to 4} {\delta\over (d-4) \delta \sigma(x)}
\Big(\int_{\cal M} d^d x'\sqrt{ g}~{E_4}(x')- \int_{\partial {\cal M}} d^{d-1} x'
\sqrt{h}~E^{\rm{(bry)}}_4(x')\Big)=\sqrt{g} {E_4}-\sqrt{h} {E^{\rm{(bry)}}_4} \ ,\\ 
\label{idk3}
&&\lim_{d\to 4} {\delta\over (d-4) \delta \sigma(x)} \int_{\partial
{\cal M}} d^{d-1} x'\sqrt{h}~ \tr\hat{K}^3(x') = \sqrt{h} \tr\hat{K}^3\
, \\ &&\lim_{d\to 4} {\delta\over (d-4) \delta \sigma(x)} \int_{\partial
{\cal M}} d^{d-1} x'\sqrt{h}~h^{\alpha\gamma}\hat{K}^{\beta\delta}W^{(d)}_{\alpha\beta\gamma\delta}
(x') = \sqrt{h} h^{\alpha\gamma}\hat{K}^{\beta\delta}W^{(4)}_{\alpha\beta\gamma\delta} 
\ . 
\ee The first identity is the consequence of the topological
invariance when the boundary term of the Euler Characteristic is
included; the last two identities reflect that they are covariant Weyl
tensors.

We find a compact expression for the boundary type-$b_1$ anomaly given
by 
\be 
\label{newexp} \tr\hat{K}^3 ={1\over 2}
\delta^{\mu\nu\lambda}_{\rho\sigma\delta} \hat{K}_\mu^\rho
\hat{K}_\nu^\sigma \hat K_\lambda^\delta \ .
\ee using three Kronecker
delta functions. In the dimensional regularization, we write 
\be
\label{exdimreg} \int_{\partial \cal{M}} d^{d-1} x \sqrt{h} \tr\hat{K}^3 =
\int_{\partial \cal{M}} \frac{ \left( \bigwedge_{j=1}^{d-1} dx^{\mu_j} \right)
}{2(d-4)!} \hat{K}_{\mu_1}^{a_1} \hat{K}_{\mu_2}^{a_2} \hat
K_{\mu_3}^{a_3} e^{a_{4}}_{\mu_{4}} \cdots e^{a_{d-1}}_{\mu_{d-1}}
\epsilon_{a_1 \cdots a_{d-1}}  \ , 
\ee 
similar to the way we express the
bulk Euler density in $d$ dimensions using vielbeins in \eqref{int}.

We will work in Gaussian normal coordinates, $x^\mu=\{x_{\perp},
x^i\}$, such that $x_{\perp} =0$ is the local function for the boundary.
The metric is given by
\be 
\label{normal} 
ds^2= dx^2_{\perp} + h_{ij}(x_{\perp},
x^i) dx^i dx^j \ . 
\ee 
Focusing on the response from varying the bulk
metric of $\cal M$, we keep the boundary metric $h_{ij}(x_{\perp}=0,
x^i)$ fixed while we perform the variation.  However, the normal
derivative of the metric variation on the boundary can be non-zero in
general.  Under the coordinate \eqref{normal}, on the boundary we
require 
\be 
\label{bry1} 
&&\delta g_{ \mu\nu}|_{\partial {\cal{M}}}=
\delta h_{ \mu\nu}(x_{\perp}=0,x^i)=0\ , \\ 
\label{bry2}
&&\partial_{\rm{n}} \delta g_{{\rm{n}} i}|_{\partial
{\cal{M}}}=\partial_{\rm{n}} \delta g_{{\rm{n}}{\rm{n}}}|_{\partial
{\cal{M}}}=\delta
{n}_\mu|_{\partial {\cal{M}}}=0 \ , \\
 \label{bry3} &&\partial_{\rm{n}} \delta
g_{ij}|_{\partial {\cal{M}}}= \partial_{\rm{n}} \delta
h_{ij}(x_{\perp},x^i)|_{\partial {\cal{M}}} \neq 0 \ . 
\ee 
The metric
variation of extrinsic curvatures on the boundary then can be written by\footnote{Relevant metric variation of curvatures can be found in the
appendix.}
\be 
\label{k} 
\delta {K_{\mu\nu}}|_{\partial {\cal{M}}} = {1\over 2} { h^\lambda_\mu h_\nu^\rho } \partial_{\rm{n}} \delta
g_{\lambda \rho}|_{\partial {\cal{M}}} \ , ~ \delta {K}|_{\partial {\cal{M}}}={1\over 2} h^{\lambda\sigma} \partial_{\rm{n}}\delta
g_{\lambda\sigma}|_{\partial {\cal{M}}} \ . 
\ee 
We will not discuss the correlation functions of the
boundary stress tensor obtained by varying the boundary metric, $h$. In
this case the contribution instead comes from the tangential fluctuation
of the metric along the boundary, and it will be independent of the bulk
stress tensor correlation functions considered here.

It is instructive to verify the identity \eqref{idk3} using 
expression \eqref{exdimreg}.  We should show 
\be 
\label{k3i}
\sqrt{h} \tr\hat{K}^3= \lim_{d\to 4} {2\over d-4} g^{\m}_{\n} {\delta \over \delta g^{\m}_{\n}}\int_{\partial \cal{M}} \frac{ \left( \bigwedge_{j=1}^{d-1}
dx^{\mu_j} \right) }{2(d-4)!} \hat{K}_{\mu_1}^{a_1}
\hat{K}_{\mu_2}^{a_2} \hat K_{\mu_3}^{a_3} e^{a_{4}}_{\mu_{4}} \cdots
e^{a_{d-1}}_{\mu_{d-1}} \epsilon_{a_1 \cdots a_{d-1}}  \ . 
\ee 
In the metric variation we receive contributions from varying the extrinsic
curvatures and also from varying the vielbeins.\footnote{We impose the boundary condition on the metric only in the physical dimensions. In the dimensional regularization scheme, the extra dimensions introduce additional vielbeins, whose variation on the boundary should be kept non-vanishing. It is the variation of these vielbeins that allows us to correctly reproduce the anomaly by varying an effective action.} 
The latter procedure
results in terms $\sim \hat K^\omega_\rho \hat K^\sigma_\eta \hat
K^\lambda_\delta \delta^{\rho\eta\delta \m}_{\omega\sigma\lambda \n}$
and identity \eqref{idk3} can be recovered by contracting $\m$ with
$\n$.  But one then
needs to show that the contribution from varying the extrinsic curvatures
with respect to metric is traceless.  We can see this is indeed the
case by writing the explicit metric expansion  as 
\be 
\label{exhk} 
\hat K_{\alpha}^{\beta} \sim \Big(K_{\alpha}^{\beta} +
{1\over 2} { h^\lambda_\alpha h^{\beta\rho} } \partial_{\rm{n}} \delta
g_{\lambda \rho}\Big) -{h^{\beta}_{\alpha}\over
{d-1}}\Big(K+ {1\over 2}
h^{\lambda\rho} \partial_{\rm{n}}\delta g_{\lambda\rho}\Big) 
= \hat K_{\alpha}^{\beta} +{1\over 2}
H^{~\lambda\beta\rho}_\alpha \partial_{\rm{n}}\delta g_{\lambda\rho}
\ , 
\ee 
where a tensor is introduced, 
\be
H^{~\lambda\beta\rho}_\alpha= { h^\lambda_\alpha h^{\beta\rho} }
-{1\over d-1}h^{\beta}_{\alpha} h^{\lambda\rho}\ , 
\ee with the
desired traceless property:
$H^{~\lambda\a\rho}_\alpha=H^{~~~\beta\lambda}_{\alpha\lambda}=0$. The
higher-order expansions of \eqref{exhk} vanish due to the boundary
condition. Therefore, all we need to include on the boundary is the
first-order part.

\subsection{Renormalization Group Flow with a Boundary}

Taking into account the boundary terms of the conformal anomaly
\eqref{4dtrace}, we would like to see how the RG equation \eqref{rge4d}
might be affected by boundary central charges. We shall focus on
potential $\mu$-dependent divergences coming from the boundary local
counter-terms.

First we discuss the boundary term of the a-type anomaly.   With the
boundary term, the fact that the Lovelock gravity has a well-defined 
variational principle in the presence of a boundary implies that there
will be no boundary contribution to the metric variation left
over. In dimensions $d=4+\epsilon$, the additional variation comes
from varying the vielbeins, which only appear through the wedge
products. Varying these vielbeins gives additional indices in the
totally antisymmetric  tensor that vanishes in the limit $d\to
4$ \footnote{For certain geometries one could obtain finite stress tensors. We will discuss them  in the next section. However, for the RG equation only the $\mu$-dependent poles are relevant.}. In short, the topological nature of Euler characteristic is still
preserved in a non-compact manifold and the RG flow remains to be
independent of the a-charge. The argument applies to any even dimensions.

The story is more subtle for the boundary Weyl invariants. Let us
first consider the boundary $b_2$-type.  In a recent paper
\cite{Solodukhin:2015eca}, it is shown that the metric variation of the
following action: 
\be
\label{solos}
 S=\int_{{\cal M}}  d^dx\sqrt{g}
~W_{\a\b\mu\nu}^n-n\int_{\partial{\cal M}}  d^{d-1}x\sqrt{h}
~P^{\a\b\mu\nu} W_{\a\b\mu\nu}^{n-1}\ , 
\ee 
does not have boundary terms containing normal derivatives of the metric variation left over.  The tensor
$P_{\alpha\beta\mu\nu}$ has the same symmetries as the
$W_{\alpha\beta\mu\nu}$ and is defined by 
\be
P_{\alpha\beta\mu\nu}=P^{(0)}_{\alpha\beta\mu\nu}-\frac{(g_{\alpha\mu}P^
{(0)}_{\beta\nu}-g_{\alpha\nu}P^{(0)}_{\beta\mu}-g_{\beta\mu}P^{(0)}_{\alpha\nu}+g_{\beta\nu}P^{(0)}_{\alpha\mu})}{d-2}
+\frac{P^{(0)}(g_{\alpha\mu}g_{\beta\nu}-g_{\alpha\nu}g_{\beta\mu})}{(d-
1)(d-2)} \ , 
\ee 
where 
\be &&P^{(0)}_{\alpha\beta\mu\nu}=n_\alpha n_\nu
K_{\beta \mu}-n_\beta n_\nu K_{\alpha \mu}- n_\alpha n_\mu K_{\beta
\nu}+ n_\beta n_\mu K_{\alpha \nu}\ ,\\ 
&&P^{(0)}_{\beta\nu}=g^{\a
\m}P^{(0)}_{\alpha\beta\mu\nu}=-n_\b n_\n K-K_{\beta\nu} \ , \\
&&P^{(0)}=g^{\beta\nu}P^{(0)}_{\beta\nu}= - 2K \ . 
\ee 
Note that
$P^\alpha_{\ \mu \alpha\nu}=0$. Under the conformal transformation,
$P_{\alpha\beta\mu\nu}\rightarrow e^{3\sigma} P_{\alpha\beta\mu\nu}$. 
To relate this formulation with the effective action of the type-B
anomaly, we take $n=2$ in \eqref{solos}.  Using $\tr(PW)=\tr (
P^{(0)}W)=4\hat{K}_{\nu\alpha} W^{\mu\nu\alpha\beta}n_\mu n_\beta $ we write 
\be 
\label{solo} 
\tilde W= {c \over 16 \pi^2} \Big( \int_{{\cal
M}}d^d x \sqrt{g}~ W^{2(d)}_{\mu\nu\lambda\rho}+8~ \int_{\partial{\cal
M}} d^{d-1} x \sqrt{h}~ h_{\mu\beta}
\hat{K}_{\nu\alpha}W^{(d)\mu\nu\b\a}\Big) \ .
\ee 
The effective action  then might be defined by multiplying the above result by the
$\mu^{\epsilon}\over \epsilon$ factor. It is interesting to notice that the coefficient
``8" of the boundary term suggests $b_2= 8c$ \cite{Solodukhin:2015eca}.

There are, however, some suspicions about this derivation that leads
to the relation $b_2= 8c$, which would mean that $b_2$ is not an independent
central charge. First of all, from the classical viewpoint, in order to
integrate the equations of motion of Weyl gravity, one needs
to specify more boundary data.  The theory of Weyl gravity might be still consistent (at the
classical level) even if on the boundary there are some normal
derivatives of the metric variation left over. (The Lovelock
gravity is special in that the action is quadratic in time derivatives.)   On the other hand, at
the quantum level, it becomes not clear that the anomaly effective
action must have a nice variational principle. In other words, one might
allow some (traceless) delta function distributions on the boundary as a
part of the contribution to the type-B anomaly induced stress tensor.

Because of these potential issues, one might adopt the following
decomposition of the type-b anomalies related
effective action: 
\be 
\label{ddc} 
\widetilde W_{B}={\mu^{\epsilon}\over
\epsilon}\Big[ {c \over 16 \pi^2} \Big(\int_{\cal M}
W^{2 (d)}_{\mu\nu\lambda\rho}+8\int_{\partial {\cal M}}
h_{\mu\beta}
\hat{K}_{\nu\alpha}W^{(d)\mu\nu\b\a}\Big)+{1\over 16 \pi^2} \int_{\partial {\cal M}} \Big( b_1 \tr\hat{K}^3+b'_2
h_{\mu\beta}
\hat{K}_{\nu\alpha}W^{(d)\mu\nu\b\a}
\Big)\Big] \ . 
\ee 
From the conformal invariance requirement,
we are allowed to add more
$h\hat{K}W$ on
the boundary with the coefficient measuring the difference
$b'_2=b_2-8c$.  Indeed, if the argument using the variational principle
is invalid, one needs another derivation to verify $b'_2$. However, we notice that the spin $0, {1\over 2}, 1$ case's free field calculations given recently in \cite{Fursaev:2015wpa} suggest the following universal result (independent of boundary conditions)
\be 
\label{b22} 
b_2'=0 \ . 
\ee 
This is the scenario we adopt in what follows.  It is still interesting, though, to find a general proof of the result \eqref{b22}  without referring to the variational method.

Let us go back to the correlation function. As the consequence of the
decomposition \eqref{ddc} and the result \eqref{b22}, the $b_2$-type boundary term 
basically plays the role of a Gibbons-Hawking-like term for the bulk
type-B anomaly effective action. Notice that \eqref{solo} applies
directly in $d=4+\epsilon$ dimensions. Therefore, this boundary
$b_2$-type  does not generate a $\mu$-dependent pole in the
three-point function, and hence the RG flow is not touched by the
$b_2$-charge.

Finally we consider the boundary $b_1$-type. We will see this type of
boundary Weyl invariant contributes to the RG equation in the vicinity of the boundary.  This central charge depends on boundary conditions; see \cite{Fursaev:2015wpa} for related heat kernel
computation.  Focusing on the $\mu$-dependent, singular contribution in
the $d\to4$ limit, we denote the correction as 
\be 
&& \label{mainresult} 
\Delta \la T_{\mu\nu}(x) T_{\sigma\rho}(y) T_{\alpha\beta}(z)
\ra^{(b_1)} = -{b_1\over 2 \pi^2} {\mu^{\epsilon}\over \epsilon}
D^{(b_1)}_{\mu\nu\sigma\rho\alpha\beta} (x,y,z) \ , 
\ee 
where 
\be
D^{(b_1)}_{\mu\nu\sigma\rho\alpha\beta}(x,y,z)=  {\delta^3
\over \delta{g^{\mu\nu}(x) \delta g^{\sigma\rho}(y) \delta g^{\a\b}(z)}}
\int_{\partial {\cal M}} d^{d-1} x \sqrt{h} \tr\hat{K}^3\ . 
\ee Using expression \eqref{exdimreg}, we obtain 
\be 
\label{check}
\lim_{d\to 4} D^{(b_1)}_{\mu\nu\sigma\rho\alpha\beta}(x,y,z)=-{3\over 8}
\delta^{\delta \omega \lambda }_{\eta \zeta \xi} H^{~~~\eta}_{\delta(\m~\n)}(x)
H^{~~~\zeta}_{\omega (\sigma~\rho)}(x) H^{~~~\xi}_{\lambda(\a~\b)}(x)  \partial_{\rm{n}}
\delta (x_{\perp}) \partial_{\rm{n}} \delta^4(x-y)\partial_{\rm{n}}
\delta^4(x-z) \ . 
\ee The normal component does not contribute and the correction only exists near the boundary.

In summary, as the generalization of \eqref{rge4d}, the RG equation of the
three-point function for $d=4$ QFTs in a flat manifold with a boundary
is given by 
\be 
\label{rg4b}
\mu {\partial\over \partial \mu} \la T_{\mu\nu}(x)
T_{\sigma\rho}(y) T_{\alpha\beta}(z) \ra =- {c\over 2 \pi^2}
\lim_{d\to 4}  D^W_{\mu\nu\sigma\rho\a\b} (x,y,z) - {b_1\over 2 \pi^2}
\lim_{d\to 4}  D^{(b_1)}_{\mu\nu\sigma\rho\alpha\beta} (x,y,z)\ .
\ee 
$D^W_{\mu\nu\sigma\rho\a\b}$ is determined by the Q-curvature via \eqref{qq} and the boundary correction is given in \eqref{check}.

\subsection{Conformal Ward Identity with a Boundary and Stress Tensor}

We next revisit the conformal Ward identity by including 
boundary anomalies. As the generalization of \eqref{ward}, we find
\be 
\label{cb} &&
\la T_{\mu}^{\mu} (x) T_{\sigma\rho}(y) T_{\a\b}
(z)\ra\nn\\ &=& 2  \Big(\delta^4(x-y)+\delta^4(z-x) \Big)\la
T_{\sigma\rho}(y)  T_{\a\b} (z) \ra+ {1\over 4 \pi^2} \Big(c
B_{\sigma\rho\a\b}(x,y,z)-aA_{\sigma\rho\a\b}(x,y,z)\Big)\nn\\ &&+ {\delta(x_{\perp}) \over 4 \pi^2} \Big(a
E_{\sigma\rho\a\b} (x,y,z)+ b_1 
b^{(1)}_{\sigma\rho\a\b}(x,y,z)+ b_2
b^{(2)}_{\sigma\rho\a\b}(x,y,z)\Big)\nn\\ 
&&-{ a\over 2 \pi^2} {\delta(x_{\perp})}\delta^{\m\n\lambda}_{\delta\eta\zeta} K^{\delta}_\mu(x) K^\eta_\nu (x) \Big(\delta_{\a\b}
h^\zeta_{(\rho} h_{\lambda |\sigma)}(x) \delta^4(x-z) \partial_{\rm{n}}\delta^4(x-y)+ \sigma \leftrightarrow \a, \rho \leftrightarrow \b, y \leftrightarrow z \Big) \nn\\
&&+{3  b_1\over 16
\pi^2}\delta(x_{\perp})\delta^{\mu\nu\lambda}_{\rho\sigma\delta}
\hat{K}_\mu^\rho(x) \hat{K}_\nu^\sigma(x) \Big({\delta_{\a\b}}
H_{\lambda(\sigma~~\rho)}^{~~~~\delta}(x)  \delta^4 (x-z)
\partial_{\rm{n}} \delta^4(x-y)+ \sigma \leftrightarrow \a, \rho
\leftrightarrow \b, y \leftrightarrow z \Big)\nn\\ 
&&+S_{\sigma\rho\a\b}(x,y,z) \la T^\m_\m(x) \ra|_{\delta_{\m\n}} \ ,
\ee  
where  
\be
\la T_{\mu}^{\mu}(x)\ra|_{\delta_{\m\n}}={\delta(x_{\perp})\over 16 \pi^2}  \Big( b_1 \tr\hat{K}^3-{8\over 3}a \delta^{\mu\nu\lambda}_{\sigma\rho\delta}
K^{\sigma}_{\mu } K^{\rho}_{\nu} K^{\delta}_{\lambda} \Big) \ ,
\ee  
is the flat limit of the conformal anomaly.
The results $A_{\sigma\lambda\rho\eta}(x,y,z)$ and
$B_{\sigma\lambda\rho\eta}(x,y,z)$ are given by \eqref{ab1} and \eqref{ab2},
respectively; the tensor $S_{\sigma\rho\a\b}$ is defined 
in \eqref{s}. We also have 
\be
E_{\sigma\rho\a\b}(x,y,z)&=&\lim_{g_{\mu\nu}\to \delta_{\mu\nu}}
{\delta^2 \over \delta{g^{\a\b}(z) \delta g^{\sigma\rho}(y)}} 
E^{\rm{(bry)}}_{4}(x)\nn\\ &=& {-4} ~
\delta^{\mu_1\mu_2\mu_3}_{\nu_1\nu_2\nu_3}~K^{\nu_1}_{\mu_1}(x)
h^{\nu_2}_{(\b}  h_{\mu_2|\a)} (x)h^{\nu_3}_{(\rho} h_{\mu_3|\sigma)}(x) \partial_{\rm{n}} \delta^4(x-y)\partial_{\rm{n}} \delta^4(x-z)\ , \\
b^{(1)}_{\sigma\rho\a\b}(x,y,z)&=&\lim_{g_{\mu\nu}\to
\delta_{\mu\nu}} {\delta^2 \over \delta{g^{\a\b}(z) \delta g^{\sigma\rho}(y)}}  \tr \hat K^3(x)\nn\\ &=& {3 \over 4}~ 
\delta^{\mu_1\mu_2\mu_3}_{\nu_1\nu_2\nu_3} \hat K_{\mu_1}^{\nu_1}
(x)H_{\mu_1(\a~~~\b)}^{~~~~~\nu_2 }(x)H_{\mu_3(\sigma~~~\rho)}^{
~~~~~\nu_3}(x)\partial_{\rm{n}} \delta^4(x-y)\partial_{\rm{n}}
\delta^4(x-z) \ ,\\
b^{(2)}_{\sigma\rho\a\b}(x,y,z)&=&\lim_{g_{\mu\nu}\to
\delta_{\mu\nu}} {\delta^2 \over \delta{g^{\a\b}(z) \delta g^{\sigma\rho}(y)}} h^{\a\gamma}
\hat{K}^{\beta\delta}W_{\alpha\beta\gamma\delta}(x)=0 \ .
\ee 
We have used the fact that when the Weyl tensor or the Riemann tensor is pulled-back on the
boundary, there are no normal derivatives acting on the metric variation left over.

Notice that there is a $\mu$-dependent pole 
from the two-point function proportional to the c-charge. As noticed by
\cite{Osborn:1993cr}, this singular behaviour can be
reproduced (or derived from definition \eqref{tt}) by requiring
compatibility between the three point function and the Ward
identity \eqref{id2}. The result is given by
\be 
\la T_{\mu\nu}(x) T_{\sigma\rho}(y) \ra^{(c)} \sim-{c\over 4\pi^2} {\mu^{\epsilon}\over \epsilon} 
\Delta^T_{\m\n\sigma\rho} \delta^4(x-y),
~~\Delta^T_{\m\n\sigma\rho}
={1\over2}(S_{\m\sigma}S_{\n\rho}+\sigma
\leftrightarrow \rho)-{1\over 3}S_{\m\n}S_{\sigma\rho} \ ,
\ee 
where $S_{\m\n}=(\partial_\m \partial_\n-\delta_{\m\n} \Box)$.  (For the contact term, see
\eqref{contact}.) 
Similarly, the boundary charge $b_1$ also contributes a pole to the
two-point function. We have
\be
\label{2bc} 
\la T_{\m\n}(x) T_{\sigma \rho}(y) \ra^{(b_1)} \sim {3 b_1\over 16\pi^2} {\mu^{\epsilon}\over \epsilon} \delta_{\a\b\gamma}^{\lambda \eta \zeta}
\hat{K}_{\lambda}^{\a}(x) H^{~~~~ \b}_{\eta (\mu~~\nu)}(x) H^{~~~~ \gamma}_{\zeta
(\sigma~~\rho)}(x)\partial_{\rm{n}}\delta (x_{\perp})\partial_{\rm{n}}\delta^4
(x-y) \ . 
\ee   The correction does not have the normal component and only exists near the boundary. 

Let us also discuss the expectation value of the stress tensor in the flat limit. Normally the stress tensor vanishes in flat space, but
here a curved boundary generating boundary Weyl anomalies can lead to 
non-vanishing stress tensors.  

The boundary $b_2$-type anomaly, as we have discussed earlier, does not touch the
resulting metric variation because of the Gibbons-Hawking
mechanism (we take $b_2=8c$). Therefore, there is no stress tensor correction by the $b_2$-charge.

Next we consider the contribution from the boundary
$b_1$-type.  We find 
\be 
\label{3db} 
\la T_{\a}^\b(x)  \ra^{(b_1)} 
&=&-{3 b_1 \over 32 \pi^2} {\mu^{\epsilon}\over \epsilon} 
\delta_{\m\n\lambda}^{\sigma\rho\eta} \hat{K}_{\sigma}^{\m}(x) \hat{K}_{\rho}^{\n}(x)  H^{~~~~
\lambda \b)}_{\eta (\a~~}(x)\partial_{\rm{n}}\delta (x_{\perp}) +{ b_1 \over 16
\pi^2} \delta{(x_{\perp})}  t_\a^\b(x)\ , 
\ee 
where the second term
comes from varying the vielbeins and is given by 
\be
t_\a^\b={\mu^{\epsilon}\over \epsilon} {1\over 2} \hat{K}_\mu^\rho
\hat{K}_\nu^\sigma \hat K_\lambda^\delta
\delta^{\mu\nu\lambda\b}_{\rho\sigma\delta\a} \ . 
\ee  
Notice that the
metric $h^{\m}_{\n}$ showing up in $\hat{K}^{\mu}_{\nu}$ contracts with
the generalized delta function and it then generates the factor
$\epsilon=d-4$, cancelling the pole. We have 
\be 
\label{lt}
 t_\a^\b
&=&{\mu^{\epsilon}\over \epsilon}{1\over 2} {K}_\mu^\rho {K}_\nu^\sigma 
K_\lambda^\delta \delta^{\mu\nu\lambda\b}_{\rho\sigma\delta\a} \nn\\ 
&&-{1\over 2}
\Big({3K\over (d-1)} {K}_\mu^\rho {K}_\nu^\sigma
\delta^{\mu\nu\b}_{\rho\sigma\a} -3({K\over (d-1)})^2 (d-3){K}_\mu^\rho
\delta^{\mu\b}_{\rho\a} +({K\over (d-1)})^3(d-3)(d-2)
\delta^\b_\a\Big)|_{d\to 4}\nn\\ 
&=&{\mu^{\epsilon}\over \epsilon}{1\over 2}
{K}_\mu^\rho {K}_\nu^\sigma  K_\lambda^\delta
\delta^{\mu\nu\lambda\b}_{\rho\sigma\delta\a}-{1\over 2} \Big(K {K}_\mu^\rho
{K}_\nu^\sigma \delta^{\mu\nu\b}_{\rho\sigma\a} -{1\over 3}K^2 {K}_\mu^\rho
\delta^{\mu\b}_{\rho\a} +{2\over 27}K^3 \delta^\b_\a\Big) \ . 
\ee 
One might argue that the first term  simply vanishes in
$(d-1)\to 3$ (the physical dimensionality of the boundary) limit, due to the additional indices in the totally-antisymmetric tensor. 
However, if the boundary geometry has a
special form of the extrinsic curvature, for example, $K^{\a}_\b= {c\over r} h^\a_\b$
where $c$ is a constant, the contraction of indices can still lead to a
finite result. In other words, the expression of the finite part of the
stress tensor depends on the structure of the boundary\footnote{Similarly, the conformal flatness condition on the
bulk geometry is implemented in \cite{Herzog:2013ed} in order to have a
finite bulk stress tensor in curved space.}.  For a ball where $\hat
K^{\a}_{\b}$ vanishes in d-dimensions, the stress tensor \eqref{3db} vanishes directly. This type of boundary anomaly does not exist for the ball-like geometry.  

The cylinder however provides a non-trivial example. The corresponding metric and extrinsic curvatures are given by
\be
\label{cylinders}
ds^2&=&dr^2+ h_{ij}(r,x^i) dx^i dx^j =dr^2+ \Big(d^2z + r^2 d^2\theta+ r^2 \sin^2\theta d^2\phi+...\Big)\ , \\
\label{cylinderk}
&&~~~~~~~K_{\m}^{\n}=  {\rm{diag}}\Big(0, 0, {1\over r}, {1\over r}, ... \Big) \ ,~ K= {d-2\over r}  \ .
\ee
It is useful to denote a delta function $\bar \delta^\m_\n=  {\rm{diag}}\Big(0,0,1,1,1,1,...\Big)$ where $r$ and $z$ components are empty. The extrinsic curvature then can be written as $K_{\m}^{\n}={1\over r} \bar \delta_\m^\n$. We consider a tube that has a radius $r_0$ so the boundary is set by $\delta(x_{\perp})=\delta (r-r_0)$.  The first term of \eqref{lt} gives
\be
\label{1t}
{K}_\mu^\rho {K}_\nu^\sigma  K_\lambda^\delta
\delta^{\mu\nu\lambda\b}_{\rho\sigma\delta\a}
&=& 
h^\b_\a \Big(K^3-3 K \tr K^2+2 \tr K^3\Big)
-3 K^\b_\a \Big( K^2-\tr K^2\Big)
+6  K^\rho_\a \Big(K K^\b_\rho-K^\sigma_\rho K^\b_\sigma\Big)\nn\\
&= & {1\over r_0^3} \Big(h^\b_\a (d-4)(d-3)(d-2)-3 \bar \delta^\b_\a (d-4)(d-3)\Big) \ , 
\ee  
which generates an overall $d-4$ factor cancelling the pole.
The second term of \eqref{lt} evaluated in $d=4$ contains the following contribution:
\be
\label{2t}
\Big(K {K}_\mu^\rho{K}_\nu^\sigma \delta^{\mu\nu\b}_{\rho\sigma\a} 
-{1\over 3}K^2 {K}_\mu^\rho
\delta^{\mu\b}_{\rho\a} 
+{2\over 27}K^3 \delta^\b_\a\Big)|_{d\to 4}
=  \Big({52\over 27 } h^\b_\a- {8\over 3} \bar \delta^\b_\a\Big) {1\over r_0^3}\ .
\ee 
From \eqref{lt}, \eqref{1t} and \eqref{2t}, by taking $d\to 4$ we obtain $t_\a^\b={1\over r_0^3} {\rm{diag}} (0, {1\over 27}, -{7\over 54},-{7\over 54})$.  
Next we consider the first piece in \eqref{3db}. 
Recalling the traceless property, we can write
\be
\delta_{\m\n\lambda}^{\sigma\rho\eta} \hat{K}_{\sigma}^{\m} \hat{K}_{\rho}^{\n}  H^{~~~~
\lambda \b)}_{\eta (\a~~}
=2\hat{K}_{\m}^{\n} \hat{K}_{\rho}^{\m}  H^{~~~~\rho \b)}_{\n (\a~~}
= -{2\over 3 r_0^{2}}\bar \delta^\m_\n   H^{~~~~\n\b)}_{\m (\a~~}\ ,
\ee  
where in the last equality we have used the cylindrical geometry.
However, we see that this non-vanishing result does not generate a $d-4$ factor, and therefore it leads to an infinite contribution to the stress tensor.
To have better behaviour, we add the following regulator:
\be
\widetilde W^{\rm{reg}}
=  {\mu^{\epsilon}\over \epsilon} c'\int_{\partial {\cal {M}}} d^3 x \sqrt{h} 
~\tr \hat K^3 \ ,
\ee
with the coefficient $c'=- {b_1\over 16 \pi^2}$ being adjusted to cancel the divergence when taking $\epsilon \to 0$.\footnote{Similar regulators are needed in order to have well-defined type-B anomaly induced stress tensors in a generally non-conformally flat background \cite{Huang:2013lhw}.}  This regulator does not touch the anomaly coefficient since it is manifestly Weyl invariant. 
We obtain the final (renormalized) stress tensor
\be
\label{rst}
\la T_{\a}^\b  \ra^{(b_1)}|_{\rm{cylinder}} 
={ b_1 \over 16\pi^2 r_0^{3}} {\rm{diag}} \Big(0, {1\over 27},- {7\over 54}, -{7\over 54}\Big) \delta{(r-r_0)} \ .
\ee    
The stress tensor contributes near the boundary and there is no r-component contribution, as expected.  Taking the trace on this stress tensor gives 
\be
\la T^\a_\a \ra^{(b_1)}|_{\rm{cylinder}}  = -{ b_1 \over 72\pi^2  r_0^{3}} \delta{(r-r_0)} \ ,
\ee 
which reproduces the Weyl anomaly evaluated for a $d=4$ cylinder where
\be
E^{\rm(bry)}_4=0,~~\tr \hat K^3=- {2\over 9 r^3_0} \ .
\ee 

We have mentioned that the Euler characteristic boundary term provides the Gibbons-Hawking mechanism so that the RG flow is untouched by the boundary counter-term $a \int_{\partial {\cal M}}E^{\rm(bry)}$. However, there can be finite contribution to the stress tensor from varying the vielbeins. 
(Varying the vielbeins on the bulk Euler characteristic counter-term gives the a-type stress tensor in curved space \cite{Herzog:2013ed} and the result vanishes in the flat limit.) Including the boundary term, the Euler characteristic gives the following contribution to the stress tensor in the flat limit
\be
\label{abt} 
\la T_{\a}^\b(x) \ra^{(a)} 
=-{a\over 6 \pi^2}{\mu^{\epsilon}\over \epsilon} \delta{(x_{\perp})} \delta_{\m\n\lambda\a}^{\sigma\rho\eta\b} K_{\sigma}^{\m}(x) K_{\rho}^{\n}(x) {K}_{\eta}^{\lambda}(x)  \ .
\ee 
Let us first consider a ball with a radius $r_0$. Starting with bulk dimensionality $d$, we have
\be
\label{ballds}
ds^2 &=& dr^2+r^2 d\theta^2 + r^2 \sin^2\theta d\phi^2+ r^2 \sin^2\theta \sin^2\phi d\psi^2+... \ , \\
\label{balldk}
&&~~~~~~~~~K^{\a}_\b= {h^\a_\b \over r} ,~~ K= {(d-1)\over r}  \ . 
\ee
We obtain a finite result
\be
\label{abt} 
\la T_{\a}^\b \ra^{(a)}_{(\rm{ball})}&=&-{a\over 6 \pi^2 r^3_0} {\mu^{\epsilon}\over \epsilon} (d-4)(d-3)(d-2)\delta (r-r_0) \delta^{\b}_{\a} \nn\\
&=&-{a\over 3 \pi^2 r^3_0} {\rm{diag}}\Big(0,1,1,1\Big) \delta (r-r_0)  \ ,
\ee  
where we have sent $d\to 4$ in the last equality. The trace of this stress tensor is
\be
\la T_{\a}^\a \ra^{(a)}_{(\rm{ball})}=-{a\over  \pi^2 r^3_0} \delta (r-r_0) \ , 
\ee 
which recovers the Weyl anomaly for a 4-ball where
\be
E^{\rm(bry)}_{4}=-{16\over r^3_0} ,~~ \hat K_{\a\b}=0 \ .
\ee 

Let us next consider a cylinder.  
Notice that $E^{\rm(bry)}_{4}$ vanishes for a 4-cylinder and the anomaly comes solely from the boundary $b_1$-charge.   One might think that, because of the vanishing $E^{\rm(bry)}_{4}$, there should not have any a-type stress tensor correction near the boundary. However, it is important to recall that the geometry  can be fixed only after working out the variation. We find
\be
\label{st2}
\la T_{\a}^\b \ra^{(a)}|_{\rm{cylinder}} 
={a\over 6 \pi^2 r_0^3} {\rm{diag}}\Big(0,-2,1,1\Big) \delta (r-r_0) \ ,
\ee  for a 4-cylinder. The result is traceless, as expected.

It is of great interest to reproduce these new stress tensors in the vicinity of a boundary whose values are determined by boundary central charges using a different approach. 
 

\section{Three-Dimensional CFTs}

\subsection{Boundary Weyl Anomalies}

Folklore has it that there is no Weyl anomaly in an odd dimension. This
statement is based on the fact that it is impossible to construct a
scalar with the correct dimension using curvatures. However, for an
odd-dimensional manifold with a boundary,  there can be Weyl anomalies
localizing on the boundary.  

For a $d=3$ manifold, the anomaly is given
by 
\be 
\label{3da} 
\langle T^\mu_\mu(x) \rangle= {a_3\over 384 \pi}\delta(x_{\perp})\oR+ {c_3\over 256 \pi}  \delta(x_{\perp}) \tr \hat K^2
\ . 
\ee  
We denote $\oR$ as the intrinsic Ricci scalar on the boundary. The trace-free
extrinsic curvature in this case is given by 
\be 
\hat K_{\mu\nu}=K_{\mu\nu}-{K\over 2} h_{\m\n}\ . 
\ee  
In this notation, we have $(a_3, c_3)=(-1, 1)$ for a conformal scalar
with the Dirichlet boundary condition and $(a_3, c_3)=(1, 1)$ for the
conformal Robin condition \cite{Branson:1995cm}. (See \cite{Solodukhin:2015eca} for a recent discussion on $d=5$ CFTs.) In this section we discuss how these boundary charges contribute to stress tensor correlation functions.

\subsection{Correlation Function and Stress Tensor}

We again focus on the correlation functions
obtained by the bulk metric variation.  We will not perturb the boundary metric, $h$, and the boundary conditions are \eqref{bry1}, \eqref{bry2} and \eqref{bry3}.   The following identity relates the boundary counter-term to the $a_3$-type anomaly:
\be 
\sqrt{h} \oR = \lim_{d\to 3} {2\over d-3} g^{\m}_{\n} {\delta \over \delta g^{\m}_{\n}} \int_{{\partial{\cal M}}}
\frac{ \left( \bigwedge_{j=1}^{d-1} dx^{\mu_j} \right) }{2(d-3)!} {{\mathring R}^{a_1
a_2}}_{~~~~\mu_1 \mu_2} e^{a_{3}}_{\mu_{3}}
\cdots e^{a_{d-1}}_{\mu_{d-1}} \epsilon_{a_1 \cdots a_{d-1}} \ . 
\ee 
The bulk metric variation does not touch the boundary Riemann curvature ${\mathring R}^{a_1a_2}_{~~~~\mu_1 \mu_2}$ which is intrinsic on ${\partial{\cal M}}$. 
The contribution comes only from varying the vielbeins. 
The $a_3$-charge does not generate a $\mu$-dependent pole, and hence the RG flows of (bulk) stress tensor correlation functions are independent from this boundary central charge. 

For the boundary Weyl invariant, $\tr \hat K^2$, we first
re-write it as 
\be 
\tr \hat K^2 = -\delta_{\a\b}^{\m\n} \hat K^\a_\m \hat  K^\b_\n \ . 
\ee 
Working now in $d=3+\epsilon$ dimensions, we have the
following identity: 
\be 
\sqrt{h}\tr\hat{K}^2= \lim_{d\to 3} {-2\over d-3}
g^{\m}_{\n} {\delta \over \delta g^{\m}_{\n} }\int_{\partial {\cal M}} \frac{
\left( \bigwedge_{j=1}^{d-1} dx^{\mu_j} \right) }{(d-3)!}
\hat{K}_{\mu_1}^{a_1} \hat{K}_{\mu_2}^{a_2}  e^{a_{3}}_{\mu_{3}} \cdots
e^{a_{d-1}}_{\mu_{d-1}} \epsilon_{a_1 \cdots a_{d-1}}  \ , 
\ee  
that relates this conformal anomaly to the boundary counter-term.  Using the
general expression \eqref{id3}, we obtain the Ward identity 
\be 
\label{3dw} 
&&\la T^\mu_\mu(x) T_{\sigma\rho}(y)T_{\a\b}(z)\ra\nn\\ 
&=&2(\delta^3(x-y)+\delta^3(x-z))\la
T_{\sigma\rho}(y) T_{\a\b} (z)\ra \nn\\ 
&& 
- {c_3\over 128 \pi} 
\delta(x_{\perp})\delta_{\lambda\delta}^{\m\n} H^{~~~~\lambda }_{\m(\sigma~~\rho)}
(x)H^{~~~~\delta}_{\n(\a~~\b)}(x) \partial_{\rm{n}} \delta^3(x-y)
\partial_{\rm{n}} \delta^3(x-z) \nn\\ 
&&-{ c_3\over 128
\pi}\delta(x_{\perp}) \delta^{\mu\n}_{\lambda\delta} \hat{K}_\mu^\lambda(x) \Big(
H_{\n(\sigma~~\rho)}^{~~~~\delta} (x){\delta_{\a\b}} \delta^3 (x-z)
\partial_{\rm{n}} \delta^3(x-y)+ \a \leftrightarrow \sigma , \b
\leftrightarrow \rho, y \leftrightarrow z \Big)\nn\\ 
&&+ S_{\a\b\sigma\rho} \la T^\m_\m(x)\ra|_{\delta_{\m\n}}
\ , 
\ee  where the flat limit of the anomaly is the same as \eqref{3da}. We have defined  $S_{\a\b\sigma\rho}$ in \eqref{s}. 

We can also compute the RG equations of the correlation functions. The
third-order metric expansion on the $c_3$-type effective action (which contains only two
$\hat K_{\m\n}$ here) does not generate a $\mu$-dependent pole, and therefore there is no correction to the RG equation of the three-point function.  For the two-point function, a relevant contribution comes
from varying the extrinsic curvatures. (Varying the vielbeins will not
give a $\mu$-dependent pole.) The RG flow is given by 
\be
\label{3rg}
\mu {\partial\over \partial \mu} \la T_{\m\n}(x) T_{\a\b}(y)\ra
={c_3\over 128 \pi} \Big( \delta_{\sigma \delta}^{\lambda \rho}
H^{~~~~\sigma}_{\lambda(\a~~\b)}(x) H^{~~~~\delta}_{\rho(\mu~~\n)}(x)\partial_{\rm{n}} \delta(x_{\perp})
\partial_{\rm{n}} \delta^3(x-y) \Big)
\ . 
\ee 
This result applies for $d=3$ QFTs in a flat manifold with a 
boundary. The correction appears only near the
boundary.

Let us next discuss stress tensors in $d=3$ flat space with a boundary. 
From the $c_3$-type anomaly, we have
\be 
\label{t3} 
\la T_{\m}^{\n}(x)\ra^{(c_3)} 
&=& {c_3\over 128 \pi}{\mu^{\epsilon}\over \epsilon} 
\delta_{\sigma\delta}^{\lambda \rho} \hat{K}_{\lambda}^{\sigma}(x)
H^{~~~~\delta\n)}_{\rho(\mu~~}(x) \partial_{\rm{n}}
\delta(x_{\perp})-{c_3\over 128 \pi} \delta(x_{\perp})  t_\m^\n(x) \ .
\ee 
Notice that $\delta_{\sigma\delta}^{\lambda \rho} h_{\lambda}^{\sigma}
H^{~~~~\delta}_{\rho(\mu~~\n)} \sim H^{~~~~\rho}_{\rho(\mu~~\n)} =0$,
and hence the $\hat{K}^{\sigma}_{\lambda}$ in the first term of \eqref{t3}
can be replaced by  ${K}^{\sigma}_{\lambda}$. The second term of \eqref{t3} 
comes from varying the vielbeins and it is given by 
\be 
\label{littlet}
t_\m^\n=
{\mu^{\epsilon}\over \epsilon}{1\over 2} \hat K^\a_\rho \hat K^\b_\sigma\delta_{\a\b\m}^{\rho\sigma\n}
= {\mu^{\epsilon}\over \epsilon}{1\over 2} K^\a_\rho  K^\b_\sigma
\delta_{\a\b\m}^{\rho\sigma\n}+{1\over 2}\Big({K^2\over 4} h_{\m}^{\n}-K 
K^\a_\rho \delta_{\a\m}^{\rho\n}\Big)\ . 
\ee 
The second contribution in \eqref{littlet} is finite and obtained from contracting $h^\a_\b$ in $\hat
K^\a_\b$ with the generalized delta function.  
For the ball-like geometry, this stress tensor
(and also this type of boundary anomaly itself) vanishes.  
For a cylinder with a radius $r_0$, using the metric and extrinsic curvatures given in \eqref{cylinders} and \eqref{cylinderk}, we have
\be
 K^\a_\rho  K^\b_\sigma\delta_{\a\b\m}^{\rho\sigma\n}
=h^\n_\m ( K^2-\tr K^2) -2(K K^\n_\m-K^\rho_\m K^\n_\rho)
&=& {h^\n_\m\over r_0^2}(d-2)(d-3)- {2\over r_0^2}\bar\delta^\n_\m(d-3) \ , \\
\Big({K^2\over 4} h_{\m}^{\n}-K K^\a_\rho \delta_{\a\m}^{\rho\n}\Big)|_{d\to 3}
&=&-{1\over r_0^2}\Big({3\over 4}h_{\m}^{\n} - \bar \delta_{\m}^{\n}\Big) \ .
\ee 
Thus, we obtain
$ t_\m^\n
={1\over 2 r_0^2}{\rm{diag}}\Big(0,{1\over 4},-{3\over 4}\Big)$.
The first term in \eqref{t3} still has the remaining $1\over \epsilon$ factor that causes the divergence.  We again add an Weyl invariant regulator
\be
\widetilde W^{\rm{reg}}
= {\mu^{\epsilon}\over \epsilon}c' \int_{\partial {\cal {M}}} d^2 x \sqrt{h} 
~\tr \hat K^2 \ ,
\ee  
with the coefficient $c'=-{ c_3\over 256\pi}$. The (renormalized) stress tensor is given by
\be
\label{rst3}
\la T_{\m}^{\n}\ra^{(c_3)}|_{\rm{cylinder}} 
= {c_3\over 256 \pi r_0^2}  {\rm{diag}} \Big(0,-{1\over 4}, {3\over 4}\Big) \delta(r-r_0) \ .
\ee  
Taking the trace on this stress tensor gives 
\be
\la T^\m_\m \ra^{(c_3)}|_{\rm{cylinder}}= {c_3\over 512 \pi r_0^2} \delta(r-r_0) \ ,
\ee which recovers the anomaly for a $d=3$ cylinder where 
\be
\oR=0,~\tr \hat K^2={1\over  2 r_0^2}  \ .
\ee

On the other hand, there can be $a_3$-type stress tensors near the boundary. 
We have
\be
\label{3da}
\la T^{\m}_{\n}(x)\ra^{(a_3)} = {a_3\over 768 \pi}{\mu^{\epsilon}\over \epsilon}  {{\mathring R}^{\lambda\sigma}}_{~~\rho\delta} \delta^{\rho\delta\m}_{\lambda\sigma\n}
\delta(x_{\perp}) = - {a_3\over 192 \pi}{\mu^{\epsilon}\over \epsilon} \mathring{E}^{\m}_\n  \delta(x_{\perp})  \ .
\ee   ($\mathring{E}^{\m}_\n$ is the boundary Einstein tensor.)
For a 3-ball with the radius $r_0$, we obtain
\be
\label{a3b}
\la T^{\m}_{\n}\ra^{(a_3)}_{\rm{ball}} 
= {a_3\over 192 \pi r_0^2}\rm{diag}\Big(0, {1\over 2}, {1\over 2}\Big)\delta(r-r_0)\ .
\ee The stress tensor trace, 
\be
\la T^{\m}_{\m}\ra^{(a_3)}_{\rm{ball}} 
=   {a_3\over 192 \pi r_0^2}\delta(r-r_0) \ , 
\ee recovers the Weyl anomaly evaluated for a 3-ball where 
\be
\oR={2\over r^2_0},~~\hat K_{\m\n}=0 \ .
\ee  
For a 3-cylinder with the radius $r_0$, we have 
\be
\label{a3c}
\la T^{\m}_{\n}\ra^{(a_3)}_{\rm{cylinder}} 
= {a_3\over 192 \pi r_0^2}\rm{diag}\Big(0, {1\over 2}, -{1\over 2}\Big)\delta(r-r_0)\ .
\ee 
This result is traceless, as expected, since the $a_3$-type anomaly vanishes for a 3-cylinder.

\section{Conclusion}

The general motivation of this paper comes from the constraints of the RG
flows in both even-and odd-dimensional QFTs in flat manifolds with a
boundary, and also comes from the universal contribution to the
entanglement entropy. It is certainly of great interest to find more
physical quantities characterized by boundary terms of conformal
anomalies. Here we investigate the stress tensor correlation functions in
flat spacetime with a generally curved boundary, focusing on the
contribution from the boundary counter-terms. In particular, in
$d=4$, we find that the conformal Ward identity is modified by boundary
central charges and the charge $b_1$ gives an additional correction to
the RG equation near the boundary. Moreover, the boundary counter-terms induce new stress tensors near the boundary. We have considered examples using a ball and a cylinder.  We also  discussed the similar story for $d=3$ CFTs, where the Weyl anomaly exists only on the boundary. It will be interesting to generalize these results to five and six dimensions.

Let us conclude by listing some questions that relate to boundary Weyl
anomalies:

(1) How do the boundary central charges modify the n-point functions of the stress tensor at non-zero separation?

(2) For four-dimensional QFTs with a boundary, (i)
is it generally true that the edge central charge satisfies
$\rm{b_{1(UV)} > b_{1 (IR)}}$?  (ii) Does the bulk $a$-charge still
satisfy monotonicity under the RG flow in the presence of a
boundary? (The boundary terms of the dilaton effective action calculated
in \cite{Herzog:2015ioa} might be useful.)

It will be also interesting to interpret these boundary Weyl anomalies in the
AdS/CFT correspondence.

\section*{Acknowledgements}

I would like to thank Chris Herzog and Kristan Jensen for reading the manuscript and for their useful comments. The work was supported in part by the NSF under Grant No. PHY13-16617.

\section{Appendix} 
\subsection{Metric Variation} 

For the convenience of the reader, here we list formulae for the metric variation of curvatures. 

Under the metric perturbation $g_{\mu\nu}\to g_{\mu\nu}+ \delta g_{\mu\nu}$, we
have 
\be 
 g^{\mu\nu}&\to& g^{\mu\nu}- g^{\mu \alpha}g^{\nu \b} \delta
g_{\alpha\beta}+g^{\mu \alpha}g^{\nu \b} g^{\lambda\rho}\delta
g_{\alpha\lambda}\delta g_{\b\rho}+\cdots \ ,\\
\sqrt{g} &\to& \sqrt{g}+ {1\over
2}\sqrt{g} g^{\mu\nu}\delta g_{\mu\nu}+\cdots \ , \\
\delta^{(n)}\Gamma^\lambda_{\mu\nu}
&=&{n\over 2} \delta^{(n-1)} (g^{\lambda \rho}) \Big(\nabla_\mu\delta
g_{\rho\nu}+\nabla_\nu \delta g_{\rho\mu}-\nabla_\rho \delta 
g_{\mu\nu}\Big) \ , \\ 
 \delta {R^\lambda_{~\mu\sigma\nu}}&=&\nabla_\sigma
\delta\Gamma^\lambda_{\mu\nu}-\nabla_\nu
\delta\Gamma^\lambda_{\mu\sigma} \ ,\\
\delta {R_{\mu\nu}}&=& {1\over 2} \Big(
\nabla^\lambda \nabla_\mu \delta g_{\lambda \nu}+\nabla^\lambda
\nabla_\nu \delta g_{\mu\lambda}-g^{\lambda\rho}\nabla_\mu \nabla_\nu
\delta g_{\lambda\rho}-\Box \delta g_{\mu\nu} \Big) \ , \\ 
\delta {R}
&=& - R^{\mu\nu} \delta g_{\mu\nu} + \nabla^\mu \Big(\nabla^\nu \delta
g_{\mu\nu}-g^{\lambda \rho}\nabla_\mu \delta g_{\lambda \rho} \Big)\ .
\ee

Defining the induced metric by $h_{\mu\nu}= g_{\mu\nu}-n_\mu n_\nu$, on
the boundary we have 
\be 
\delta n_\mu &=& {1\over 2}n_\mu n^\lambda
n^\nu \delta g_{\lambda\nu} \ , \\
\delta {K_{\mu\nu}}&=& \delta
(h_\mu^\lambda h_\nu^\rho \nabla_\lambda n_\rho)\\
 &=&  {n^\lambda
n^\rho \delta g_{\lambda\rho} K_{\mu\nu}\over 2} +\delta g_{\lambda
\rho} n^\rho \Big(n_\mu K^\lambda_\nu+n_\nu K^\lambda_\mu \Big)- {
h^\lambda_\mu h_\nu^\rho n^\alpha\over 2} \Big(\nabla_\lambda \delta
g_{\alpha\rho}+ \nabla_\rho \delta g_{\lambda\alpha}-\nabla_\alpha
\delta g_{\lambda \rho}\Big) \ , \nn\\
\delta {K}&=&-{1\over 2}K^{\mu\nu}
\delta g_{\mu\nu}-{1\over 2} n^\rho\Big( \nabla^\lambda \delta
g_{\rho\lambda}- g^{\lambda\sigma} \nabla_\rho \delta
g_{\lambda\sigma}\Big) - {1\over 2} \oD^\mu \Big(h^{\lambda}_{\mu}
n^\rho \delta g_{\lambda\rho}\Big) \ ,
\ee where $\oD^\mu$ denotes the covariant derivative on the boundary.

\subsection{Two-dimensional CFTs with a Boundary}

For a compact $d=2$ manifold, the Weyl anomaly is given by
\be
\label{2dtrace}
\la T^\m_\m(x) \ra= {a\over 2\pi} R = {c\over 24\pi} R  \ . 
\ee  The central charge notation c is more common in the literature. 
One has 
\be
\int \sqrt{g} d^2x R=2\pi \chi \ ,
\ee where $\chi$ is the Euler number. Using the expansion on the
Ricci scalar around flat space, one obtains the Ward identity 
\be 
\la T^\m_\m(x) T_{\sigma \lambda}(y)\ra &=& - {c\over 12\pi} (\partial_\sigma\partial_\lambda
-\delta_{\sigma\lambda} \Box) \delta^2 (x-y) \ . 
\ee   

In the presence of a boundary one has
\be 
\la T^\m_\m(x)\ra= {c\over 24\pi}
\Big(R+2K\delta{(x_{\perp})}\Big) \ , 
\ee   The Ward identity is modified
by the boundary term and we obtain
\be 
\la T^\m_\m (x) T_{\sigma
\lambda}(y)\ra
&=&- {c\over 12\pi} \Big((\partial_\sigma \partial_\lambda
-\delta_{\sigma\lambda} \Box) \delta^2 (x-y)-
\delta{(x_{\perp})} h_{\sigma\lambda} \partial_{\rm{n}}\delta^2 (x-y)
\Big) \nn\\
&&+ \delta_{\sigma\lambda} \delta^2(x-y) \la T^\m_\m(x)\ra|_{\delta_{\m\n}}+2\delta^2(x-y)\la T_{\sigma\lambda} (x)\ra|_{\delta_{\m\n}} \ . 
\ee  
The anomaly in the flat limit is given by 
\be
\label{2daf}
\la T^\m_\m \ra|_{\delta_{\m\n}} = {c\over 12\pi} K \delta{(x_{\perp})} \ .
\ee   
We define the vacuum stress tensor of a plane to vanish. For a disk with a radius $r_0$, we have 
\be
\la T_{\a}^{\b} \ra|_{\delta_{\m\n}} ={c\over 12\pi}{\mu^{\epsilon}\over \epsilon}   K^\mu_\nu \delta^{\n\b}_{\m\a}\delta{(r-r_0)}
={c\over 12\pi r_0}  {\rm{diag}}\Big(0,1\Big)\delta{(r-r_0)}
\ee  
(Now we are working in $d=2+\epsilon$ dimensions.) Only the angular component is non-vanishing. The trace of this stress tensor reproduces the anomaly in the flat limit.
Notice that in $d=2$ curved space, the bulk stress tensor has to be expressed in terms of the Weyl factor $\sigma(x)$ defined via $g_{\m\n}=e^{2\sigma} \delta_{\m\n}$, because the Einstein tensor vanishes in $d=2$. See \cite{Herzog:2015ioa} for more discussions on $d=2$  CFTs.

\end{document}